# CREMP: Conformer-rotamer ensembles of macrocyclic peptides for machine learning

Colin A. Grambow[1] ✉, Hayley Weir[1], Christian N. Cunningham[2], Tommaso Biancalani[3] & Kangway V. Chuang[1] ✉

Computational and machine learning approaches to model the conformational landscape of macrocyclic peptides have the potential to enable rational design and optimization. However, accurate, fast, and scalable methods for modeling macrocycle geometries remain elusive. Recent deep learning approaches have significantly accelerated protein structure prediction and the generation of small-molecule conformational ensembles, yet similar progress has not been made for macrocyclic peptides due to their unique properties. Here, we introduce CREMP, a resource generated for the rapid development and evaluation of machine learning models for macrocyclic peptides. CREMP contains 36,198 unique macrocyclic peptides and their high-quality structural ensembles generated using the Conformer-Rotamer Ensemble Sampling Tool (CREST). Altogether, this new dataset contains nearly 31.3 million unique macrocycle geometries, each annotated with energies derived from semi-empirical extended tight-binding (xTB) DFT calculations. Additionally, we include 3,258 macrocycles with reported passive permeability data to couple conformational ensembles to experiment. We anticipate that this dataset will enable the development of machine learning models that can improve peptide design and optimization for novel therapeutics.

## Background & Summary

**Introduction.** Macrocyclic peptides are an emerging class of therapeutics in drug discovery[1,2]. Recent advances in affinity selection and display technologies have enabled ultra-large scale screening libraries that identify high-affinity and selective binders for challenging-to-drug proteins[3,4]. Importantly, cyclic peptides occupy a unique biophysical space between small molecules and proteins, and cyclization imparts key properties such as increased proteolytic stability and conformational rigidity[5]. The resulting flexible-yet-constrained geometries enable macrocycles to bind to shallow protein surfaces and disrupt protein-protein interactions[6]. Intriguingly, the same dynamic conformational behavior that drives high-affinity binding also drives complex chameleonic properties such as permeability[7,8]. Despite their therapeutic potential, this critical conformational behavior is intrinsically challenging to model—accurately modeling geometry and the immense conformational space is difficult to scale. Machine learning methods that can efficiently approximate high-level computational approaches have the potential to enable rational design, yet there currently exist no large-scale datasets of macrocycle structures.

**Current Challenges and Approaches.** Efficient and accurate conformer generation for macrocycles is challenging due to many factors, including their vast structural diversity, stereochemistry, number of rotatable bonds, and complex intramolecular interactions[7]. Furthermore, the challenges of modeling the vast conformational landscape is compounded by complex solvent-dependent effects[9]. Computational approaches to macrocycle conformer generation broadly leverage both heuristics- and physics-based algorithms to accurately model and sample molecular geometries. For example, both the open-source cheminformatics library RDKit[10–13] and commercial conformer generation program OpenEye OMEGA Macrocycle[14,15] combine distance geometry algorithms with molecular force fields[16] for diverse macrocycle sampling. Alternatively, low-mode[17,18] and Monte-Carlo[19] search methods have been found to be effective for sampling when combined with molecular dynamics as implemented in Schrödinger's MacroModel[20] and Prime MCS[21] packages. A critical limitation in the application of these methods to virtual screening is their high computational cost that limits their scalability. A single macrocyclic peptide, with its large number of rotatable bonds and cyclic constraints, readily takes $10^3 - 10^6 \times$ more compute compared to a drug-like small molecule, with even more dramatic computation times when molecular dynamics approaches are used with explicit solvation[9,22].

[1]Prescient Design, Genentech, 1 DNA Way, South San Francisco, CA, 94080, USA. [2]Department of Peptide Therapeutics, Genentech, 1 DNA Way, South San Francisco, CA, 94080, USA. [3]Biology Research | Development, Genentech, 1 DNA Way, South San Francisco, CA, 94080, USA. ✉e-mail: grambow.colin@gene.com; chuang.kangway@gene.com

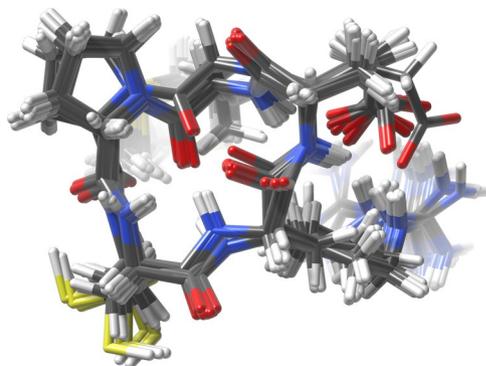

**Fig. 1** A small macrocyclic peptide ensemble generated with CREST.

| Residues | Molecules | Conformers | | | | | |
|---|---|---|---|---|---|---|---|
| | | Count | Mean | Median | Std. Dev. | Min. | Max. |
| 4 | 17,842 | 12,205,128 | 684 | 508 | 677 | 1 | 12,268 |
| 5 | 13,644 | 14,134,609 | 1,036 | 825 | 824 | 6 | 8,486 |
| 6 | 4,712 | 4,921,068 | 1,044 | 879 | 764 | 28 | 5,619 |
| Total | 36,198 | 31,260,805 | 864 | 656 | 768 | 1 | 12,268 |

**Table 1.** Dataset statistics for CREMP.

**Datasets to Enable Machine Learning.** Recently, machine learning approaches have made significant progress in small-molecule energy prediction[23–27], conformer generation[28–33], and protein structure prediction[34–39]. A critical and enabling aspect of these machine learning approaches has been access to large, high-quality training datasets. Experimentally, structural databases such as the Cambridge Structural Database (CSD)[40] and Protein Data Bank (PDB)[41] provide $10^5$ to $10^6$ training examples. Similarly, computational datasets such as QM9[42,43] and GEOM[44] have inspired machine learning research to achieve DFT-level accuracy at a fraction of the computational cost[24–26]. Unfortunately, extensive datasets on macrocycles remain scarce with only dozens to hundreds of available experimental structures. Computational datasets such as PEPCONF[45] and SPICE[46] datasets have recently begun to provide high-quality conformations at high levels of theory, but are rich in linear peptides; the total number of macrocycles remains insufficient for training machine learning models.

Here, we present a novel computational dataset of macrocyclic peptides, CREMP (**C**onformer-**R**otamer **E**nsembles of **M**acrocyclic **P**eptides), generated with the Conformer-Rotamer Ensemble Sampling Tool (CREST) package[47]. CREST leverages an iterative metadynamics algorithm with a genetic structure-crossing algorithm to explore diverse geometries, and leverages semiempirical quantum mechanical GFNn-xTB[48] calculations to provide better geometries and energy estimates than classical force fields. Initial explorations have demonstrated the potential for CREST to generate diverse macrocycle ensembles that recapitulate key intramolecular hydrogen bonds and the feasibility for ring interconversion, which is further supported by our Technical Validation below. Example conformers from a representative ensemble generated by CREST are illustrated in Fig. 1. We generate a unique set of 36,198 representative 4-, 5-, and 6-mer homodetic cyclic peptides and perform CREST simulations to generate extensive conformer ensembles in chloroform[49], providing nearly 31.3 million unique conformers with energy annotations. Chloroform was chosen as the implicit solvation model to approximate the nonpolar environment within cell membranes and facilitate modeling of membrane permeability. Although modeling aqueous environments is critical, especially for docking simulations in biological systems, achieving high-level accuracy requires more involved explicit solvent simulations, which are prohibitive for a dataset of this size within our computational allocation[22]. In total, this dataset constitutes 3.9 million CPU hours of compute. Summary statistics of the entire dataset are shown in Table 1.

Additionally, building upon the recently published CycPeptMPDB database[50], which contains experimental passive membrane permeability data compiled from a variety of literature sources but lacks structural information, we augment a selection of 3,258 6-, 7-, and 10-mer cyclic peptides with their CREST-computed conformer ensembles, yielding nearly 8.7 million additional unique conformers in this CREMP-CycPeptMPDB dataset. To our knowledge, this constitutes the first large-scale dataset containing both complete 3D-structural ensembles along with permeability annotations. We believe that such data will enable enhanced insight into the relationship between 3D structure and permeability. Summary statistics for these data are shown in Table 2.

| Residues | Molecules | Conformers | | | | | |
|---|---|---|---|---|---|---|---|
| | | Count | Mean | Median | Std. Dev. | Min. | Max. |
| 6 | 1,511 | 4,313,071 | 2,854 | 2,382 | 1,919 | 31 | 11,073 |
| 7 | 1,732 | 4,321,378 | 2,495 | 2,249 | 1,387 | 208 | 9,348 |
| 10 | 15 | 49,157 | 3,277 | 3,626 | 1,757 | 116 | 7,015 |
| Total | 3,258 | 8,683,606 | 2,665 | 2,297 | 1,666 | 31 | 11,073 |

**Table 2.** Dataset statistics for CREMP-CycPeptMPDB macrocyclic peptides.

We make these datasets publicly available to provide the research community with a valuable data resource to train and develop machine learning models, and anticipate that it will find broad utility for learning models capable of predicting macrocyclic peptide properties, such as binding affinity, stability, and permeability, among others. We hope the insights gained from these models can be used for rational design of novel macrocycles with improved therapeutic potential. We anticipate that this work will not only accelerate the development of new macrocyclic therapeutics but also contribute to a deeper understanding of the factors governing their conformational behavior, paving the way for more efficient computational approaches in the future.

## Methods

**Overview.** In this study, we used CREST to generate diverse conformer ensembles for a diverse set of macrocyclic peptides, with the aim of developing a representative dataset that captures the complexity of their conformational landscape. CREST was chosen due to its ability to efficiently explore the conformational space while providing a balance between computational cost and accuracy compared to force field-based methods.

**Sequence Identities and Processing.** Many factors contribute to the diverse conformational shapes of macrocyclic peptides, including ring size, side chains, stereochemistry, and chemical modifications. To gain coverage of a diverse set of molecules, we began with a previously enumerated set of cyclic tetrapeptides from Chan *et al.*[51] and supplemented with additional tetra-, penta-, and hexapeptides, limiting our set to exclusively homodetic peptides. Macrocycle sequences were randomly sampled across three key parameters including 1) side chain - the canonical 20 side chains, 2) stereochemistry - we used both L- and D-amino acids, and 3) *N*-methylation. The resulting set covered a total of 36,198 unique sequences. All macrocycle sequences were converted to their corresponding canonical SMILES in RDKit to check for duplicates and simple dataset statistics were collected using RDKit (number of atoms, etc.).

Additionally, 3,258 sequences were selected from the CycPeptMPDB[50] database spanning both homo- and heterodetic hexa-, hepta-, and decapeptides.

**RDKit Conformer Generation.** CREST performs best when initialized with a structure that is sufficiently close to the lowest-energy conformer. To generate initial geometries, we used RDKit ETKDGv3[11,12] with macrocycle torsion preferences to generate conformers using `EmbedMultipleConfs` with up to 5,000 conformers (`numConfs=5000`) and random coordinate initialization (`useRandomCoords=True`), which has been shown to be beneficial for generating macrocycle geometries[12]. Conformers were subsequently optimized using MMFF94[16] as implemented in RDKit and sorted by energy. Finally, the sorted conformers were filtered based on heavy-atom RMSD with a threshold of 0.5 Å.

**Optimization with xTB.** CREST recommends initialization with a structure optimized at the same level of theory as used for the energy evaluations during the simulation. Therefore, for each molecule, we reoptimized the 1,000 lowest-energy conformers from RDKit/MMFF94 using GFN2-xTB[48] and the ALPB solvent model in chloroform[49]. The resulting lowest-energy optimized structure was used as the single input to CREST.

**CREST Simulation.** The single xTB-optimized structure was used as input to CREST using solvation in chloroform and default arguments, which include an energy window of 6.0 kcal/mol, an RMSD threshold of 0.125 Å, and an energy threshold between conformers of 0.05 kcal/mol. Each calculation used 14 cores in order to efficiently parallelize across 14 metadynamics runs, and took an average of 2.3 hours for 4-mers, 5.8 hours for 5-mers, and 13.9 hours for 6-mers.

**Graph Reidentification.** It is possible for low-energy chemical reactions to occur during the CREST simulation, e.g., intramolecular proton transfers, which may produce ensembles that contain different constitutional isomers. We used OpenEye's OEChem library to reidentify chemical graphs for each conformer in an ensemble and only retained ensembles where all conformers have the same chemical graph.

## Data Records

We make the CREMP dataset available online[52] at https://doi.org/10.5281/zenodo.7931444 with examples on how to load and analyze data in our public GitHub repository. We provide the data in two available formats, either as Python pickle files, which provide quick read access with RDKit version 2022.09.5 or later, and as text-based SDF files with associated metadata in JSON format. Each file is named based on its linearized amino acid sequence, with residues separated by periods, using standard one-letter codes with lowercase letters representing D-amino acids and `Me` prefixes representing *N*-methylated amino acids. The linearized sequences are in no particular order, e.g., `C.R.E.M.P` and `R.E.M.P.C` correspond to the same peptide macrocycle (no duplicated macrocycles are present in the dataset as outlined previously). The filename extensions are `.pickle`, `.sdf`, and `.json`.

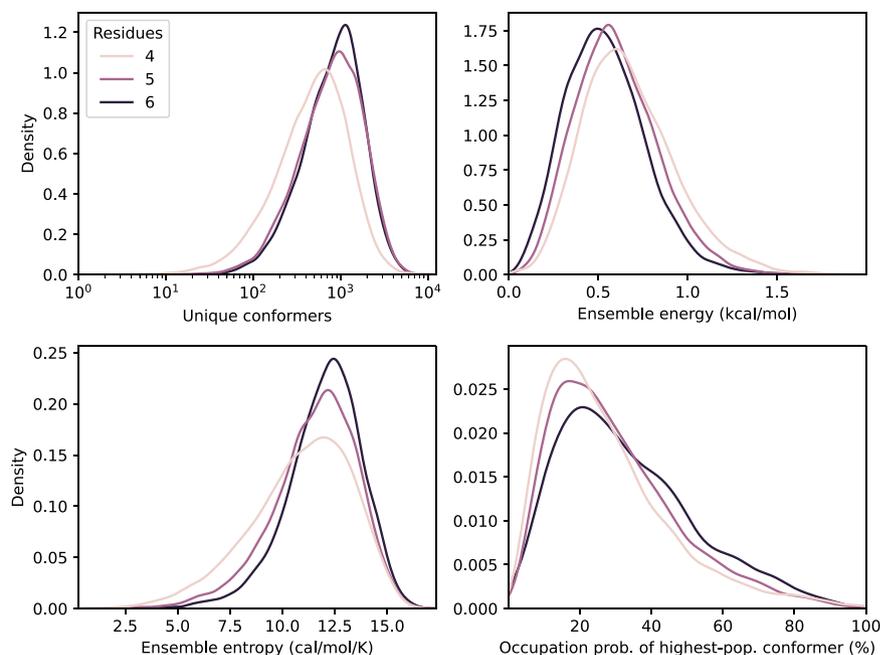

**Fig. 2** Distributions of CREST-ensemble quantities.

Each file in `pickle` contains a Python dictionary with amino acid sequence, SMILES, CREST metadata (energy, entropy, etc.), and a single RDKit `rdkit.Chem.Mol` object containing all conformers. All conformers of each molecule were validated to be the same structural isomer, and all files were compressed into a single archive. In `sdf_and_json`, each individual SDF contains all conformers, each associated with its own JSON file that contains CREST metadata. Similarly, all are compressed into another single archive. A single summary CSV is also provided containing `sequence`, `smiles`, `num_monomers`, `num_atoms`, `num_heavy_atoms`, along with the CREST metadata `totalconfs`, `uniqueconfs`, `lowestenergy`, `poplowestpct`, `temperature`, `ensembleenergy`, `ensembleentropy`, and `ensemblefreeenergy`. The number of unique conformers with different 3D structures is given by `uniqueconfs`, while `totalconfs` includes the number of rotamers in addition.

We also make the CREMP-CycPeptMPDB dataset available[53] at https://doi.org/10.5281/zenodo.10798261, which contains the conformer ensembles of a subset of macrocyclic peptides available in the CycPeptMPDB database[50] containing cell membrane permeability annotations. The format of this data record is identical to that of CREMP.

### Technical Validation

In addition to manually verifying the generated ensembles and designing the graph reidentification step so that all conformers in each ensemble have the same chemical graph, we analyzed the generated structures and performed additional characterization of CREST to validate its effectiveness in the space of macrocyclic peptides.

**Analysis of Structures.**  We performed a systematic evaluation of the data generated to visualize and quantify the distributions of conformers within CREMP in order to characterize the conformational and structural diversity of the generated ensembles. The overall dataset statistics are shown in Table 1. The ensembles are diverse in size ranging from ones as small as a single low-energy conformer to a single ensemble with more than 12,000 conformers in the 6 kcal/mol energy window.

Figure 2 visualizes the distributions of ensemble size, energy, entropy, and the occupation probability of the highest-population conformer. The occupation probability distributions demonstrate that it is generally not sufficient to only consider the lowest-energy conformer and that including many low-lying states may be required.

Figure 3 illustrates the conformer ensembles of three different macrocyclic peptides using UMAP[54]. Each UMAP plot is generated by computing the sine and cosine of each cycle torsion and projecting them into two dimensions across the entire conformer ensemble for each peptide. The distinct plots and several clusters within each plot demonstrate the conformational diversity of backbones within each macrocycle and across peptides. Additionally, each conformer is colored based on its number of internal hydrogen bonds, its radius of gyration, and its energy relative to the lowest-energy conformer in the ensemble. The different values of these properties for each cluster further illustrate the diversity of backbone conformations and the variation within each cluster. In particular, each backbone conformational cluster tends to be associated with a unique number of internal hydrogen bonds, which may stabilize a particular backbone conformation.

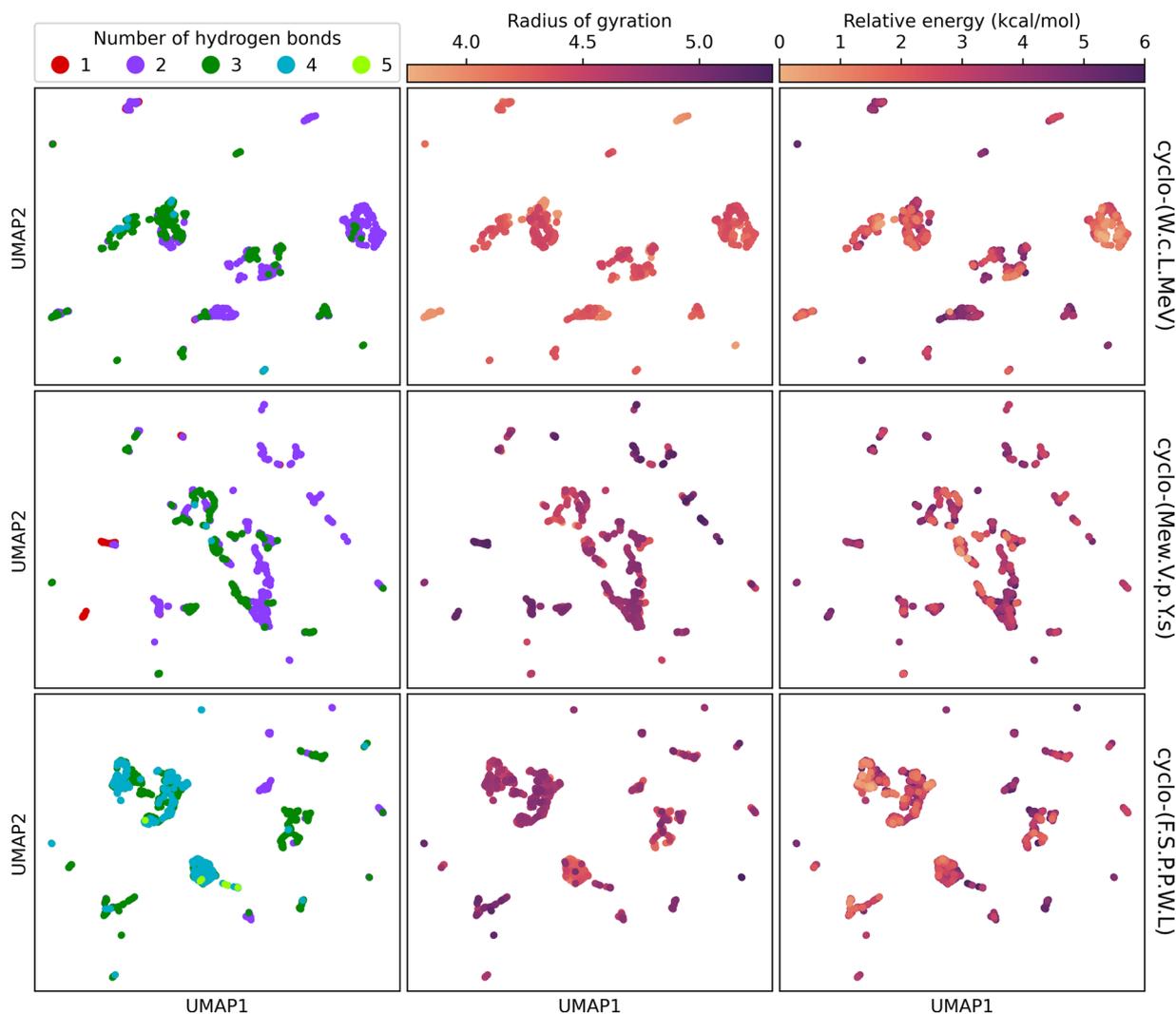

**Fig. 3** Two-dimensional projections of sine- and cosine-transformed cycle torsions generated using UMAP and colored by 3D properties. Each point represents a unique conformer. Each peptide macrocycle has a distinct conformer ensemble characterized by several backbone conformations corresponding to the clusters on the UMAP plots.

**Structural Validation with NMR Ensembles.** CREST has already been shown to recover relevant conformers of some macrocyclic peptides[47], but such assessment has been limited in scope due to low availability of experimental data on structural ensembles of macrocycles. Moreover, we expect both the experimental and computational methods to have a strong effect on the structures. In particular, the continuum solvent approximation ALPB[49] is likely not sufficient for strongly polar solvents, particularly when they are hydrogen-bond donors.

To investigate this, we performed a limited analysis on NMR ensembles available in the PDB. In order to characterize the quality of the CREST ensembles computed for the PDB structures, we compute **Mat**ching (MAT) scores[30]. For each molecule:

$$\text{MAT-RMSD}\left(\mathbb{S}_{\text{CREST}}, \mathbb{S}_{\text{PDB}}\right) = \frac{1}{|\mathbb{S}_{\text{PDB}}|} \sum_{\mathbf{R}' \in \mathbb{S}_{\text{PDB}}} \min_{\mathbf{R} \in \mathbb{S}_{\text{CREST}}} \text{RMSD}\left(\mathbf{R}, \mathbf{R}'\right) \quad (1)$$

where $\mathbb{S}_{\text{CREST}}$ is the set of conformations generated by CREST and $\mathbb{S}_{\text{PDB}}$ is the set of experimental NMR structures. **R** denotes a conformation and RMSD is computed using heavy atoms only. Intuitively, smaller MAT-RMSD scores correspond to CREST ensembles that contain more realistic conformers and are able to recover the relevant NMR conformers. Similarly, we define a MAT score for the ring torsion fingerprint deviation (rTFD):

$$\text{MAT-rTFD}\left(\mathbb{S}_{\text{CREST}}, \mathbb{S}_{\text{PDB}}\right) = \frac{1}{|\mathbb{S}_{\text{PDB}}|} \sum_{\mathbf{R}' \in \mathbb{S}_{\text{PDB}}} \min_{\mathbf{R} \in \mathbb{S}_{\text{CREST}}} \text{rTFD}\left(\mathbf{R}, \mathbf{R}'\right) \quad (2)$$

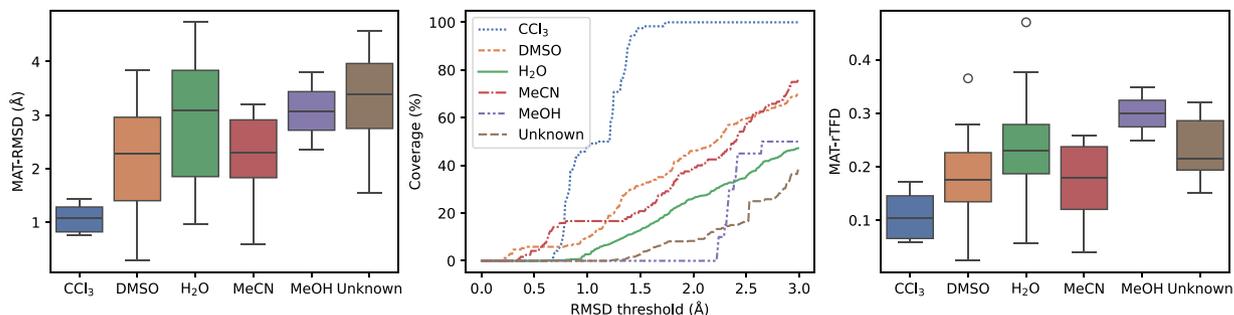

**Fig. 4** Matching (MAT) scores for RMSD (left), mean coverage at different RMSD thresholds (middle), and MAT for rTFD (right) between CREST ensembles and NMR ensembles of macrocyclic peptides obtained from the PDB in different solvents.

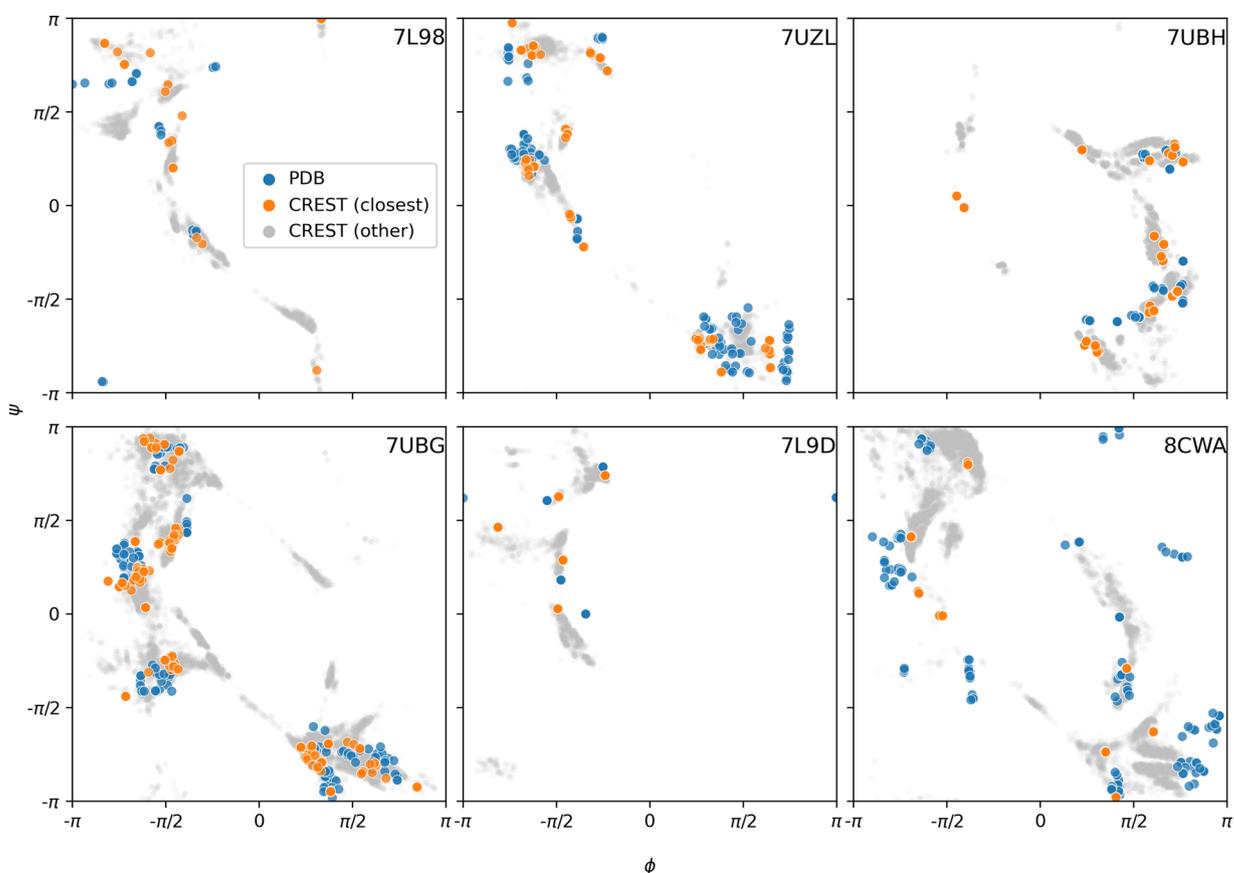

**Fig. 5** Ramachandran plots comparing CREST ensembles to NMR ensembles of macrocyclic peptides obtained from the PDB.

where rTFD quantifies how well the torsion angles in the macrocycle match between two conformations **R** and **R**′:[12]

$$\text{rTFD}\,(\mathbf{R},\,\mathbf{R}') \,=\, \frac{1}{N_{\text{torsions}}} \sum_{i=1}^{N_{\text{torsions}}} \frac{\left|w\left(\tau_i(\mathbf{R})\,-\,\tau_i(\mathbf{R}')\right)\right|}{\pi} \tag{3}$$

Here, $\tau_i(\mathbf{R})$ computes the $i$-th torsion angle for conformation **R** and $w(\cdot)$ wraps its argument around the cyclic domain $[-\pi, \pi)$ on which torsion angles are defined. The $\pi$ in the denominator normalizes each deviation so that rTFD lies in [0, 1]. Smaller values of MAT-rTFD correspond to CREST ensembles that accurately reproduce the ring torsions of the reference NMR data. We also compute a coverage metric, which measures the fraction of correctly generated conformers within an RMSD threshold of $\delta_{\text{RMSD}}$[30]. In other words, the coverage metric quantifies how many experimental conformers are recovered with CREST. It is given by:

$$\text{Coverage } (\mathbb{S}_{\text{CREST}}, \mathbb{S}_{\text{PDB}}) = \frac{1}{|\mathbb{S}_{\text{PDB}}|} \, |\{\mathbf{R}' \in \mathbb{S}_{\text{PDB}} : \exists \mathbf{R} \in \mathbb{S}_{\text{CREST}}, \; \text{RMSD } (\mathbf{R}, \mathbf{R}') \leq \delta_{\text{RMSD}}\}|  \quad (4)$$

Figure 4 illustrates the range of MAT-RMSD and MAT-rTFD scores computed between CREST and NMR ensembles in various solvents as well as the coverage over a range of RMSD thresholds. While the performance of CREST is relatively poor in polar solvents, especially water, likely due to the use of implicit solvation, it excels at generating high-quality ensembles in chloroform with MAT-RMSD ~1 Å and MAT-rTFD ~0.1. Moreover, the coverage plot shows that all experimental conformers in chloroform can be recovered within a threshold of ~1.5 Å.

To provide more fine-grained insight into the ensembles in chloroform, we plotted the conformers from six available PDB structures along with the corresponding CREST ensembles on Ramachandran plots[55] in Fig. 5 and highlighted the CREST conformers closest to the PDB conformers based on rTFD, i.e., those that result from the minimization in Eq. (2). The high-density regions of the CREST ensemble overlap with the locations of the PDB conformers, thereby further demonstrating the ability of CREST to generate high-quality ensembles in chloroform.

### Usage Notes

Data are provided in two main formats: binary pickle files and text-based SDF and JSON files. SDF and JSON are provided because they are human-readable, software-agnostic, and guarantee maximum interoperability. SDF files can also be read by every major cheminformatics program, which makes them especially suitable for users that are not as familiar with the Python programming environment. The pickle files are provided because they are much faster to load and directly yield RDKit molecules that can be processed in machine learning and cheminformatics workflows. However, pickle files have the downside that they require specific Python packages and sometimes specific versions of those. For the data published here, the Python environment at https://github.com/Genentech/cremp provides details about the required packages and their versions and provides instructions for loading the data. RDKit is a major dependency and version 2022.09.5 or newer is required in order to load the data successfully.

### Code availability

Our workflow is built on several open-source toolkits including the RDKit[10], xTB[48], and CREST[47]. xTB and CREST are both freely available (https://github.com/grimme-lab/xtb/releases and https://github.com/grimme-lab/crest/releases). OpenEye Applications and Toolkits was used for graph reidentification. We provide code with instructions on how to load the data and perform simple CREST runs at https://github.com/Genentech/cremp.

### References
1. Driggers, E. M., Hale, S. P., Lee, J. & Terrett, N. K. The exploration of macrocycles for drug discovery–an underexploited structural class. *Nat. Rev. Drug Discov.* **7**, 608–624 (2008).
2. Muttenthaler, M., King, G. F., Adams, D. J. & Alewood, P. F. Trends in peptide drug discovery. *Nat. Rev. Drug Discov.* **20**, 309–325 (2021).
3. Huang, Y., Wiedmann, M. M. & Suga, H. RNA display methods for the discovery of bioactive macrocycles. *Chem. Rev.* **119**, 10360–10391 (2019).
4. Vinogradov, A. A., Yin, Y. & Suga, H. Macrocyclic peptides as drug candidates: Recent progress and remaining challenges. *J. Am. Chem. Soc.* **141**, 4167–4181 (2019).
5. Shinbara, K., Liu, W., van Neer, R. H. P., Katoh, T. & Suga, H. Methodologies for backbone macrocyclic peptide synthesis compatible with screening technologies. *Front. Chem.* **8**, 447 (2020).
6. Villar, E. A. *et al*. How proteins bind macrocycles. *Nat. Chem. Biol.* **10**, 723–731 (2014).
7. Whitty, A. *et al*. Quantifying the chameleonic properties of macrocycles and other high-molecular-weight drugs. *Drug Discov. Today* **21**, 712–717 (2016).
8. Bhardwaj, G. *et al*. Accurate de novo design of membrane-traversing macrocycles. *Cell* **185**, 3520–3532.e26 (2022).
9. Linker, S. M. *et al*. Lessons for oral bioavailability: How conformationally flexible cyclic peptides enter and cross lipid membranes. *J. Med. Chem.* **66**, 2773–2788 (2023).
10. Landrum, G. RDKit: Open-source cheminformatics (2006).
11. Riniker, S. & Landrum, G. A. Better informed distance geometry: Using what we know to improve conformation generation. *J. Chem. Inf. Model.* **55**, 2562–2574 (2015).
12. Wang, S., Witek, J., Landrum, G. A. & Riniker, S. Improving conformer generation for small rings and macrocycles based on distance geometry and experimental torsional-angle preferences. *J. Chem. Inf. Model.* **60**, 2044–2058 (2020).
13. Wang, S. *et al*. Incorporating NOE-Derived distances in conformer generation of cyclic peptides with distance geometry. *J. Chem. Inf. Model.* **62**, 472–485 (2022).
14. Hawkins, P. C. D., Skillman, A. G., Warren, G. L., Ellingson, B. A. & Stahl, M. T. Conformer generation with OMEGA: algorithm and validation using high quality structures from the protein databank and cambridge structural database. *J. Chem. Inf. Model.* **50**, 572–584 (2010).
15. Hawkins, P. C. D. & Nicholls, A. Conformer generation with OMEGA: learning from the data set and the analysis of failures. *J. Chem. Inf. Model.* **52**, 2919–2936 (2012).
16. Halgren, T. A. Merck molecular force field. v. extension of MMFF94 using experimental data, additional computational data, and empirical rules. *J. Comput. Chem.* **17**, 616–641 (1996).
17. Kolossváry, I. & Guida, W. C. Low mode search. an efficient, automated computational method for conformational analysis: Application to cyclic and acyclic alkanes and cyclic peptides. *J. Am. Chem. Soc.* **118**, 5011–5019 (1996).
18. Kolossváry, I. & Guida, W. C. Low-mode conformational search elucidated: Application to C39H80 and flexible docking of 9-deazaguanine inhibitors into PNP. *J. Comput. Chem.* **20**, 1671–1684 (1999).
19. Chang, G., Guida, W. C. & Still, W. C. An internal-coordinate monte carlo method for searching conformational space. *J. Am. Chem. Soc.* **111**, 4379–4386 (1989).

20. Watts, K. S., Dalal, P., Tebben, A. J., Cheney, D. L. & Shelley, J. C. Macrocycle conformational sampling with MacroModel. *J. Chem. Inf. Model.* **54**, 2680–2696 (2014).
21. Sindhikara, D. *et al.* Improving accuracy, diversity, and speed with prime macrocycle conformational sampling. *J. Chem. Inf. Model.* **57**, 1881–1894 (2017).
22. Damjanovic, J., Miao, J., Huang, H. & Lin, Y.-S. Elucidating solution structures of cyclic peptides using molecular dynamics simulations. *Chem. Rev.* **121**, 2292–2324 (2021).
23. Gilmer, J., Schoenholz, S. S., Riley, P. F., Vinyals, O. & Dahl, G. E. Neural message passing for quantum chemistry. In Precup, D. & Teh, Y. W. (eds.) *Proceedings of the 34th International Conference on Machine Learning*, vol. 70 of *Proceedings of Machine Learning Research*, 1263–1272 (PMLR, 2017).
24. Smith, J. S., Isayev, O. & Roitberg, A. E. ANI-1: an extensible neural network potential with DFT accuracy at force field computational cost. *Chem. Sci.* **8**, 3192–3203 (2017).
25. Schütt, K. *et al.* Schnet: A continuous-filter convolutional neural network for modeling quantum interactions. In Guyon, I. *et al.* (eds.) *Advances in Neural Information Processing Systems*, vol. 30 (Curran Associates, Inc., 2017).
26. Gasteiger, J., Groß, J. & Günnemann, S. Directional message passing for molecular graphs. In *International Conference on Learning Representations* (2020).
27. Liu, Y. *et al.* Spherical message passing for 3d molecular graphs. In *International Conference on Learning Representations* (2022).
28. Mansimov, E., Mahmood, O., Kang, S. & Cho, K. Molecular geometry prediction using a deep generative graph neural network. *Sci. Rep.* **9**, 20381 (2019).
29. Simm, G. & Hernandez-Lobato, J. M. A generative model for molecular distance geometry. In III, H. D. & Singh, A. (eds.) *Proceedings of the 37th International Conference on Machine Learning*, vol. 119 of *Proceedings of Machine Learning Research*, 8949–8958 (PMLR, 2020).
30. Xu, M., Luo, S., Bengio, Y., Peng, J. & Tang, J. Learning neural generative dynamics for molecular conformation generation. In *International Conference on Learning Representations* (2021).
31. Xu, M. *et al.* Geodiff: A geometric diffusion model for molecular conformation generation. In *International Conference on Learning Representations* (2022).
32. Stärk, H., Ganea, O., Pattanaik, L., Barzilay, R. & Jaakkola, T. EquiBind: Geometric deep learning for drug binding structure prediction. In Chaudhuri, K. *et al.* (eds.) *Proceedings of the 39th International Conference on Machine Learning*, vol. 162 of *Proceedings of Machine Learning Research*, 20503–20521 (PMLR, 2022).
33. Jing, B., Corso, G., Chang, J., Barzilay, R. & Jaakkola, T. Torsional diffusion for molecular conformer generation. In Koyejo, S. *et al.* (eds.) *Advances in Neural Information Processing Systems*, vol. 35, 24240–24253 (Curran Associates, Inc., 2022).
34. Jumper, J. *et al.* Highly accurate protein structure prediction with AlphaFold. *Nature* **596**, 583–589 (2021).
35. Baek, M. *et al.* Accurate prediction of protein structures and interactions using a three-track neural network. *Science* **373**, 871–876 (2021).
36. Wu, R. *et al.* High-resolution de novo structure prediction from primary sequence (2022).
37. Anand, N. & Achim, T. Protein structure and sequence generation with equivariant denoising diffusion probabilistic models (2022).
38. Yim, J. *et al.* SE(3) diffusion model with application to protein backbone generation (2023) 2302.02277.
39. Wu, K. E. *et al.* Protein structure generation via folding diffusion. *Nat. Commun.* **15**, 1059 (2024).
40. Groom, C. R., Bruno, I. J., Lightfoot, M. P. & Ward, S. C. The cambridge structural database. *Acta Crystallogr B Struct Sci Cryst Eng Mater* **72**, 171–179 (2016).
41. Berman, H. M. *et al.* The protein data bank. *Nucleic Acids Res.* **28**, 235–242 (2000).
42. Ruddigkeit, L., van Deursen, R., Blum, L. C. & Reymond, J.-L. Enumeration of 166 billion organic small molecules in the chemical universe database GDB-17. *J. Chem. Inf. Model.* **52**, 2864–2875 (2012).
43. Ramakrishnan, R., Dral, P. O., Rupp, M. & von Lilienfeld, O. A. Quantum chemistry structures and properties of 134 kilo molecules. *Scientific Data* **1** (2014).
44. Axelrod, S. & Gómez-Bombarelli, R. GEOM, energy-annotated molecular conformations for property prediction and molecular generation. *Sci Data* **9**, 185 (2022).
45. Prasad, V. K., Otero-de-la Roza, A. & DiLabio, G. A. PEPCONF, a diverse data set of peptide conformational energies. *Sci Data* **6**, 180310 (2019).
46. Eastman, P. *et al.* SPICE, a dataset of drug-like molecules and peptides for training machine learning potentials. *Sci Data* **10**, 11 (2023).
47. Pracht, P., Bohle, F. & Grimme, S. Automated exploration of the low-energy chemical space with fast quantum chemical methods. *Phys. Chem. Chem. Phys.* **22**, 7169–7192 (2020).
48. Bannwarth, C., Ehlert, S. & Grimme, S. GFN2-xTB-an accurate and broadly parametrized self-consistent tight-binding quantum chemical method with multipole electrostatics and density-dependent dispersion contributions. *J. Chem. Theory Comput.* **15**, 1652–1671 (2019).
49. Ehlert, S., Stahn, M., Spicher, S. & Grimme, S. Robust and efficient implicit solvation model for fast semiempirical methods. *J. Chem. Theory Comput.* **17**, 4250–4261 (2021).
50. Li, J. *et al.* CycPeptMPDB: A Comprehensive Database of Membrane Permeability of Cyclic Peptides. *Journal of Chemical Information and Modeling.* https://doi.org/10.1021/acs.jcim.2c01573 (2023).
51. Chan, L., Morris, G. M. & Hutchison, G. R. Understanding conformational entropy in small molecules. *J. Chem. Theory Comput.* **17**, 2099–2106 (2021).
52. Grambow, C. A., Weir, H., Cunningham, C. N., Biancalani, T. & Chuang, K. V. CREMP: Conformer-Rotamer Ensembles of Macrocyclic Peptides for Machine Learning. *Zenodo* https://doi.org/10.5281/zenodo.7931444 (2023).
53. Grambow, C. A., Weir, H., Cunningham, C. N., Biancalani, T. & Chuang, K. V. CREMP-CycPeptMPDB: Conformer-Rotamer Ensembles of Macrocyclic Peptides with Permeability Annotations. *Zenodo* https://doi.org/10.5281/zenodo.10798261 (2024).
54. McInnes, L., Healy, J. & Melville, J. Umap: Uniform manifold approximation and projection for dimension reduction (2020). 1802.03426.
55. Ramachandran, G. N. & Sasisekharan, V. Conformation of polypeptides and proteins. *Adv. Protein Chem.* **23**, 283–438 (1968).

## Acknowledgements

We gratefully acknowledge Benjamin Sellers for helpful discussions on earlier versions of dataset generation. We also thank members of the Departments of Biology Research and AI Development, Peptide Therapeutics, and Discovery Chemistry at Genentech Research and Early Development for insightful discussions.

## Author contributions

C.A.G., T.B., and K.V.C. conceived the project. C.A.G. and H.W. performed the calculations and computational analysis. C.C. provided technical and scientific input on macrocyclic peptides. All authors provided feedback on the analyses. C.A.G. and K.V.C. wrote the manuscript with input and revisions from all authors.

### Competing interests
All authors are employees of Genentech, Inc. and shareholders of Roche.

### Additional information
**Correspondence** and requests for materials should be addressed to C.A.G. or K.V.C.

**Reprints and permissions information** is available at www.nature.com/reprints.

**Publisher's note** Springer Nature remains neutral with regard to jurisdictional claims in published maps and institutional affiliations.

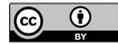 **Open Access** This article is licensed under a Creative Commons Attribution 4.0 International License, which permits use, sharing, adaptation, distribution and reproduction in any medium or format, as long as you give appropriate credit to the original author(s) and the source, provide a link to the Creative Commons license, and indicate if changes were made. The images or other third party material in this article are included in the article's Creative Commons license, unless indicated otherwise in a credit line to the material. If material is not included in the article's Creative Commons license and your intended use is not permitted by statutory regulation or exceeds the permitted use, you will need to obtain permission directly from the copyright holder. To view a copy of this license, visit http://creativecommons.org/licenses/by/4.0/.

© The Author(s) 2024